\documentclass[preprint,11pt,authoryear]{elsarticle}

\usepackage{lineno,hyperref}
\usepackage{amssymb}
\usepackage{booktabs}
\usepackage{makecell}
\usepackage{amsmath}
\usepackage{url}
\usepackage{subfigure}
\usepackage{xcolor}

\usepackage{tikz}

\usepackage{multirow}
\usepackage{booktabs}
\usepackage{pgfplots}
\pgfplotsset{compat=newest}
\usetikzlibrary{pgfplots.groupplots}
\usepgfplotslibrary{groupplots}
\usepgfplotslibrary{ternary}
\usepackage{pgfplotstable}
\usepackage{multirow}
\usetikzlibrary{arrows}
\usepgflibrary{arrows}
\usetikzlibrary{tikzmark,overlay-beamer-styles}
\usetikzlibrary{arrows.meta, positioning, matrix}

\usepackage[margin=0.94in]{geometry}

\usepackage[section]{placeins}

\journal{Simulation Modelling Practice and Theory}

\begin{document}

\begin{frontmatter}
	
		
		
		\title{Opinion dynamics and mutual influence with LLM~agents through dialog simulation} 
		
		\author[spbu]{Yulong~He}
		\author[itmo]{Dutao~Zhang}
		\author[itmo]{Sergey~Kovalchuk}
		\author[fusionbrain]{Pengyi~Li}
		\author[spbu]{Artem~Sedakov\corref{mycorrespondingauthor}}
		\cortext[mycorrespondingauthor]{Corresponding author}
		\ead{a.sedakov@spbu.ru}
		
		\affiliation[spbu]{organization={Saint Petersburg State University},
			addressline={7/9 Universitetskaya nab.},
			city={Saint Petersburg},
			postcode={199034},
			country={Russia}}
		
		\affiliation[itmo]{organization={ITMO University},
			addressline={Kronverkskiy Prospekt, 49},
			city={St. Petersburg},
			postcode={197101},
			country={Russia}}
		
		\affiliation[fusionbrain]{organization={FusionBrain Lab},
			addressline={Presnenskaya nab., 6c2},
			city={Moscow},
			postcode={123317},
			country={Russia}}
		\begin{abstract}
			A fundamental challenge in opinion dynamics research is the scarcity of real-world longitudinal opinion data, which complicates the validation of theoretical models. To address this, we propose a novel simulation framework using large language model (LLM) agents in structured multi-round dialogs. Each agent's dialog history is iteratively updated with its own previously stated opinions and those of others analogous to the classical DeGroot model. Furthermore, by retaining each agent's initial opinion throughout the dialog, we simulate anchoring effects consistent with the Friedkin--Johnsen model of opinion dynamics. Our framework thus bridges classical opinion dynamics models and modern multi-agent LLM systems, providing a scalable tool for simulating and analyzing opinion formation when real-world data is limited or inaccessible.
		\end{abstract}

		
		
		
	\end{frontmatter}

\section{Introduction}

Opinion dynamics research seeks to understand how individual attitudes evolve through social interaction and how these micro-level processes give rise to collective patterns such as consensus, polarization, and fragmentation. Classical models, such as DeGroot's (DG) iterative averaging framework and the Friedkin--Johnsen (FJ) model of anchored beliefs (see \cite{degroot1974reaching, friedkin1990social}), have provided elegant mathematical formulations to capture these dynamics. Over the past decades, these foundational models have inspired a substantial body of theoretical developments, including bounded-confidence dynamics that capture nonlinear and local interactions (\cite{deffuant2000mixing, hegselmann2002opinion}), consensus processes over switching and time-varying networks (\cite{jadbabaie2003coordination, olfati2004consensus}), and heterogeneous-influence formulations that account for structural variability and stubborn agents (\cite{proskurnikov2017tutorial}).
		
A key obstacle in advancing these theories lies in the limited availability of empirical longitudinal data. Most empirical studies rely on cross-sectional surveys (e.g. national opinion polls), short-term laboratory experiments, or large-scale digital trace data from social media platforms (see \cite{min2022multi, gong2025modeling, zareer2025survey}). While useful, these data sources have notable limitations: surveys lack temporal granularity, experiments often involve small and artificial settings, social media data are noisy and difficult to map to well-defined network structures. People's opinions are often expressed in words, and these surveys are simplified into scales, which results in the loss of a lot of information. To overcome these constraints, researchers used modern NLP techniques to convert text into embeddings (\cite{he2025social}) or sentiment analysis (\cite{he2025opinion}), methods that can convert text into numbers. However, these methods still have limitations. On social media platforms such as Weibo, people's opinions are not a closed system. This system is constantly affected by external factors, such as emergencies, news, and platform intervention.
		
The emergence of LLMs represents a significant shift in computational social science and natural language processing (\cite{brown2020language}). Initially designed for general-purpose text understanding and generation, LLMs have grown dramatically in scale and capability, enabling them to capture complex linguistic patterns, contextual nuances, and even subtle social cues. As these models have evolved from early transformer-based architectures (\cite{vaswani2017attention}) to today's multi-billion-parameter systems, they have been increasingly applied to tasks that require reasoning over long contexts, understanding human-like discourse, and modeling interaction dynamics in online and offline environments. This progress opens opportunities to complement traditional opinion dynamics models, allowing researchers to simulate and analyze social behaviors at a linguistic and cognitive level that was previously infeasible.
		
Recent work has begun to exploit LLMs as agents in multi-agent simulations, enabling richer forms of coordination, negotiation, and collective decision-making through natural language interaction. With generative agent architectures, \cite{park2023generative} and \cite{chuang2024simulating} demonstrate how memory, reflection, and planning can give rise to believable individual and emergent social behaviors. \cite{qiu2020structured} further highlight the importance of capturing interaction structure in multi-party conversations using structured dialog modeling approaches. In applied multi-agent systems, LLM-based agents effectively collaborate through language in complex tasks such as software development and recommender systems, as discussed in \cite{qian2024chatdev} and \cite{shu2024rah}. 
		
More directly related to opinion dynamics, recent studies investigate how opinions evolve within populations of interacting LLM agents. \cite{cisneros2025biases} analyzes opinion exchange among LLMs in funding-allocation scenarios, identifying systematic biases toward consensus, caution, and ethical considerations, as well as the role of memory and opinion representation in shaping collective outcomes. \cite{wu2025hidden} contrast explicit coordination mechanisms with implicit consensus formation via in-context learning, demonstrating that preserving partial diversity can enhance robustness and adaptability in dynamic environments. \cite{song2025greater} explore how multi-agent LLM systems can exert social influence on humans, revealing stronger opinion shifts when users interact with multiple agents rather than a single one. \cite{zhang2025llm} apply an LLM-based diffusion simulation in social networks under a binary opinion state model, whereas \cite{nudo2026generative} employ LLMs to simulate political discourse when agents adopt a trinary opinion state. \cite{piao2025emergence} simulate LLM discussions incorporating randomness, with a focus on polarization in the absence of any well-established opinion dynamics model. \cite{coppolillo2025engagement} demonstrate the validity of a reinforcement learning approach for LLM-based simulation.
		
At a broader level, \cite{gao2024large} survey LLM-empowered agent-based modeling and position these approaches as a rapidly evolving paradigm for simulating social, economic, and hybrid systems, while \cite{curvo2025traitors}, using recent testbeds such as The Traitors, highlights the potential of multi-agent LLM simulations to study trust, deception, and strategic communication under asymmetric information.
		
Building on this line of research, our work positions LLM-based dialog simulations as a novel tool for opinion dynamics, bridging the gap between classical mathematical models and the empirical challenges of real-world data collection. By integrating sentiment analysis to translate agent-generated text into numerical opinion values, our framework ensures compatibility with traditional opinion dynamics formalisms while preserving linguistic nuance. In this paper, we introduce a novel simulation framework that leverages LLM agents engaged in structured multi-round dialogs. Each agent iteratively updates its opinions based on prior conversational exchanges, thereby implementing influence dynamics analogous to the DG or FJ models. By retaining each agent's initial stance throughout the interaction, the framework naturally reproduces anchoring effects consistent with the FJ model. Moreover, we provide fine-grained control over inter-agent connectivity, enabling simulations on arbitrary network topologies such as fully connected graphs or small-world structures. These customizable links govern which agents can observe and influence one another, allowing researchers to investigate network-aware opinion dynamics, including consensus formation, influence hierarchies, and polarization. We employ sentiment analysis, a well-established natural language processing task that identifies the emotional or evaluative polarity expressed in the text. In addition, we compared the applicability of different LLMs to the viewpoint dynamics model, as well as the degree of mutual trust among the various LLMs.
		
Thus, our contribution is threefold. First, we bridge classical opinion dynamics theory with modern multi-agent LLM systems, offering a scalable and expressive tool for simulating social influence processes grounded in natural language. Second, we provide an experimental testbed for studying opinion formation when real-world longitudinal data are scarce or inaccessible. Finally, we assess the suitability of various LLMs for simulating opinion dynamics, identify the most effective model for this purpose, and compare patterns of mutual trust emerging among different models. Together, these advances open new opportunities for both theoretical validation and applied research in computational social science.

\section{Methodology}\label{sec3}

Our experimental framework is designed to study opinion formation and evolution in a dynamic multi-agent setting. Using OpenAI's API, we build a multi-agent opinion exchange platform where multiple LLMs act agents and take part in iterative rounds of discussion. During each round, agents generate or update their opinions by combining system prompts with the available dialog history. This iterative process mimics discussion social groups or networks and provides insight into how opinions evolve in such a dynamic environment. We describe the components of our methodology below.\footnote{Code available at: https://github.com/hreyulog/llm\_opinion\_dynamic.git}

\subsection{Initial opinion generation}\label{sec3-1}

We begin with an intentionally fictional and neutral discussion topic: \texttt{``What kind of person is Noah?''}, referred to as \texttt{\{topic\}} below. This choice helps to reduce bias from pre-existing world knowledge that may be embedded in LLMs. As a result, the model's responses are driven primarily by the interaction rules rather than artifacts from their training data. It also allows us to observe natural patterns in discussion and opinion change. Further, each agent is (randomly) assigned a personal stance on a 5-point scale: \emph{strongly positive}, \emph{positive}, \emph{neutral}, \emph{negative}, \emph{strongly negative}. This exogenous attribute remains unchanged throughout the discussion and is used only once to shape the agents' initial opinions on the chosen topic, reflecting personality traits commonly found in people, even when discussing neutral topics.
		
All agents generate their initial opinions using the same system prompt, adjusted to reflect their assigned stance:
		
\begin{quote}
	\ttfamily
	You, along with other individuals like yourself, are participating in a dis\-cus\-sion about the topic \{topic\} on an online microblogging platform such as Twitter or Weibo. The discussion takes place over several ``rounds''.
			
	In each round, every participant posts a short message expressing their views on \{topic\}. After reading the messages posted by others, you will compose a new short message on the same topic. You may choose to maintain your original stance, modify it partially, or completely change your view based on what others have said.
			
	During the entire discussion, you are not allowed to communicate with anyone outside the group or leave the discussion space to seek external information about \{topic\}. 
			
	You must rely solely on your initial understanding of the topic and the pers\-pec\-ti\-ves shared by other participants to form or revise your opinion.
			
	Keep in mind: You should act as a real person who can form highly objective opinions. Politeness is not required.
\end{quote}
		
For the first round of opinion generation, we use a uniform starter prompt to ensure consistent response structure across agents.
	
\begin{quote}
	\ttfamily
	Please compose a short message on the topic, clearly expressing your stance as \{stance\}.
			
	Topic: \{topic\}
			
	Follow the format exactly as shown below to create a new short message less than 100 words on \{topic\}.
			
	Your message:
\end{quote}
		
This ensures uniformity in expression length, format, and stance polarity, establishing a reproducible baseline for tracking opinion evolution across agents and conditions.
		
In all subsequent rounds, we update the user prompt in the dialog history, allowing agents to incorporate responses generated by other agents in the current round and, if needed, revise their opinions accordingly. This is achieved using the following turn prompt:
		
\begin{quote}
	\ttfamily
	Here is what others have said: \{others\_opinions\}
			
	Combine others' messages with your previous opinions. Follow the format exactly as shown below to create a new short message less than 100 words on \{topic\}.
	
	Your message:
\end{quote}

\subsection{Modeling social influence}\label{sec3-2}

We formalize opinion updating mechanisms using two classical models from social network theory. First, under the DG model, each agent's opinion in the next round is a weighted average of its peers' current opinions, that is, $x(t+1) = W x(t)$, where $t \in \{1,\ldots,T\}$ indexes the discussion round, $x(t) \in \mathbb{R}^n$ is the opinion vector whose entries represent the opinions of all $n$ agents, and $W$ is a given row-stochastic influence matrix. The entries of $W$ represent inter-agent trust, that is, the weights that agents assign to their peers' current opinions. In our implementation, an agent's prompt at any non-initial round includes: (i)~the system prompt (instructions), (ii)~the agent's own opinion in the previous round, and (iii)~the current opinions of other agents, weighted according to the corresponding row of the influence matrix. This enables simulation of diverse network topologies, such as fully connected or small-world structures.
		
The DG model, however, assumes that agents share their initial beliefs only in the starting round of the discussion and never attach to them in the future once the dialog has started. To address this, we adopt the FJ extension of the DG~model $x(t+1) = S W x(t) + (I-S) x(1)$, where the given diagonal matrix $S$, with entries between~0 and~1, reflects agent-specific susceptibility to interpersonal influence, and $x(1)$ is the agents' initial opinion vector. Therefore, we include the latter in the agent's prompt context as an additional element of the dialog history, enabling the FJ~model to capture heterogeneity in stubbornness and belief persistence.

\subsection{Analytical pipeline}\label{sec3-3}

To systematically analyze the trajectories of agent opinions, we first employ sentiment analysis on the LLM-generated outputs. Each outputs is mapped to an affective score, enabling the construction of time-series representations of opinions evolution. These scores are obtained using Twitter-RoBERTa\footnote{https://huggingface.co/cardiffnlp/twitter-roberta-base-sentiment.} (\cite{barbieri2020tweeteval}). The training corpus is collected from the Twitter Academic API, covering the period from 2018 to 2021 and updated on a seasonal (quarterly) basis. The final model is trained on a total of 124 million tweets. In downstream task evaluations, it outperforms the vast majority of state-of-the-art models.
		
Having time-series representations of agents' opinions, we next feed them into a model of social influence to estimate its parameters: the influence matrix $W$ and, if the FJ~model is selected, the matrix $S$. In the following section, we discuss model variants where these parameters are estimated using ordinary least squares (see \cite{he2025opinion}). For the DG~model, the OLS estimate $\widehat{W}$ of the influence matrix $W$ solves $\widehat{W} = \arg\min\limits_W \sum_{t = 1}^{T-1} ||x(t+1) - W x(t)||^2$, subject to $W$ being row stochastic. For the FJ~model, the OLS estimates $\widehat{W}$ and $\widehat{S}$ of $W$ and $S$ are obtained by solving $(\widehat{W}, \widehat{S}) = \arg\min\limits_{W,S}\sum_{t = 1}^{T-1} ||x(t+1) - S W x(t) - (I-S) x(1)||^2$, subject to $W$ being row stochastic and $S$ being diagonal with entries in $[0, 1]$.
		
\begin{figure}[h!]
	\centering
	\includegraphics[width=0.5\columnwidth]{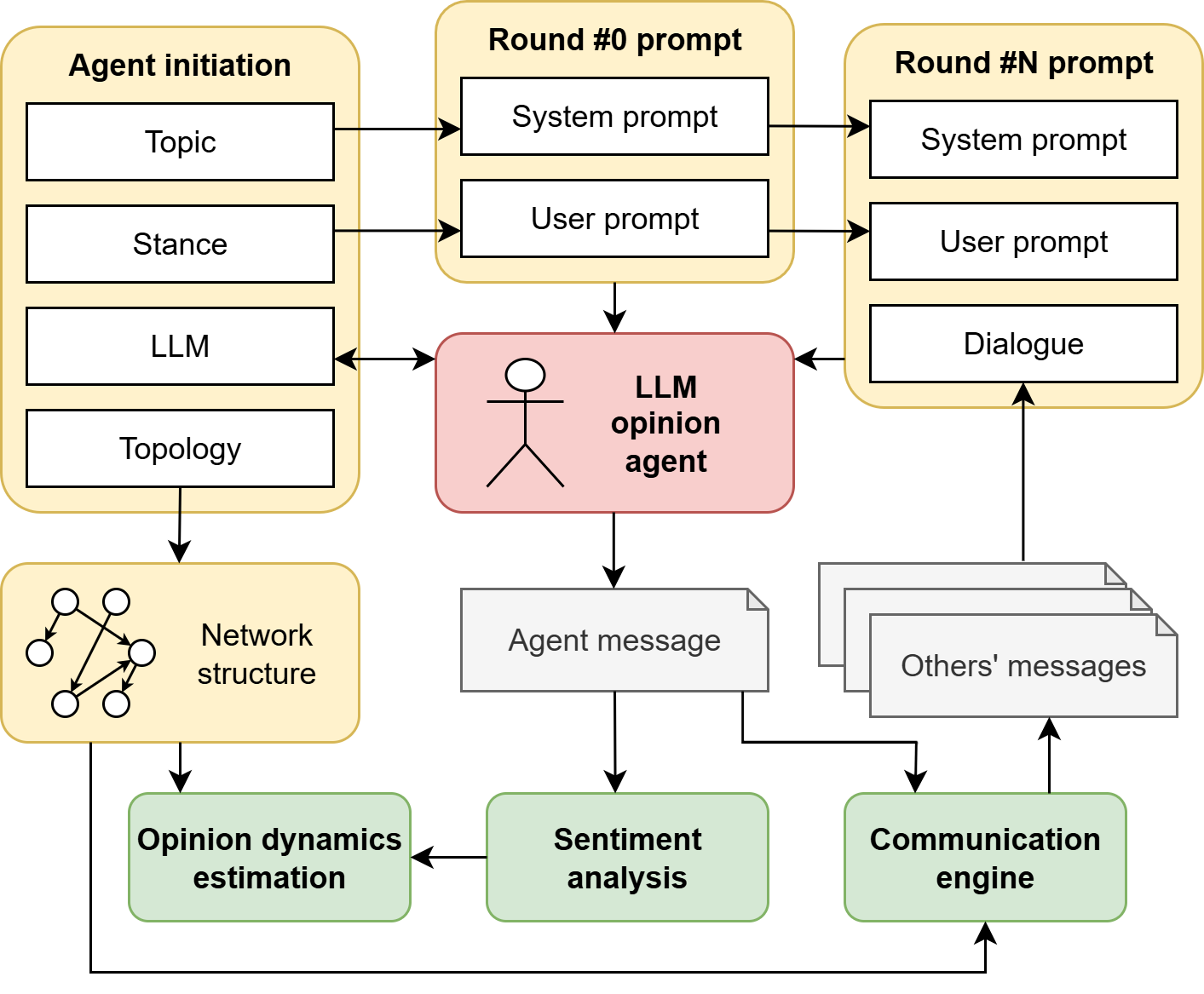}
	\caption{Overview of the multi-agent LLM opinion dynamics simulation framework.}
	\label{fig:llm_opinion}
\end{figure}

Fig.~\ref{fig:llm_opinion} shows the pipeline, which includes initial opinion generation and the iterative discussion process based on the model of social influence, implemented through the starter and turn prompts.

\section{Experiments and analysis}\label{sec4}

Simulation experiments are conducted under both DG and FJ frameworks, using homogeneous and heterogeneous agent configurations. We selected the following five LLMs to simulate group discussions:
		
\begin{itemize}
	\item \emph{GPT-4o mini (8B)} (\cite{openai2024gpt4}): A cost effective small model developed by OpenAI, based on the GPT-4o architecture. It supports up to 128K input tokens and achieves 82\% on the MMLU benchmark, offering low latency, and handling both text and vision modalities.
			
	\item \emph{Qwen2.5 (72B)} (\cite{qwen2025qwen25}): The flagship dense model of the Qwen2.5 series from Alibaba Cloud. It supports a context length of 131,072 tokens and a maximum generation length of 8,192 tokens, and excels in complex tasks such as programming and mathematics. 
			
	\item \emph{Llama~3.3 (70B)} (\cite{grattafiori2024llama3}): A multilingual, instruction tuned model developed by Meta. It is optimized for dialog and incorporates an improved Transformer architecture trained with supervised fine tuning and reinforcement learning from human feedback.
		
	\item \emph{Mistral Large (123B)} (\cite{jiang2023mistral7b}): A high performance model by Mistral AI, designed for long context reasoning and achieving state of the art results across language understanding and generation benchmarks.
			
	\item \emph{DeepSeek~V3 (671B)} (\cite{deepseekai2025deepseekv3}): A large scale model by DeepSeek~AI, focused on efficiency and deployment at scale. While its exact parameter count is not publicly disclosed, it demonstrates strong performance on reasoning and generation tasks.
\end{itemize}

\subsection{Homogeneous LLM agents}\label{sec4-1}

We begin with a homogeneous case of five identical LLM agents who only differ in their initial stance parameters, each assigned a distinct value. After simulating 20 rounds of group discussion, quantifying text responses, and estimating the model parameters, we summarize the results in Table~\ref{tab-homogeneous}. The table provides goodness of fit results for the same set of LLM agents, represented by the sum of residuals, which reflects the degree of alignment between simulated opinion trajectories and the model's updating mechanism. Additionally, the table reports the average estimates of diagonal weights $\widehat{W}_{ii}$ and $\widehat{S}_{ii}$, that is, $\mathrm{tr}(\widehat{W})/n$ and $\mathrm{tr}(\widehat{S})/n$, respectively. The former serves as a proxy for the self-trust index of agents in both DG and FJ models, whereas the latter applies only to the FJ model and measures average susceptibility to interpersonal influence.

\begin{table*}[t]
	\centering
	\footnotesize
	\caption{Goodness of fit, self-trust and susceptibility estimates for selected LLM agents under models of social influence.}
	\label{tab-homogeneous}
	\begin{tabular}{@{}l *{2}{c} *{3}{c}@{}}
		\toprule
		\multirow{2}{*}{LLM agents} 
		& \multicolumn{2}{c}{DG model} 
		& \multicolumn{3}{c}{FJ model} \\
		\cmidrule(lr){2-3} \cmidrule(l){4-6}
		& Sum of residuals & Average $\widehat{W}_{ii}$ 
		& Sum of residuals & Average $\widehat{W}_{ii}$ & Average $\widehat{S}_{ii}$ \\
		\midrule
		GPT-4o mini & 0.3276 & 0.2365 & 0.5672 & 0.2670 & 0.8226 \\
		Qwen2.5     & 0.1070 & 0.3395 & 0.2767 & 0.1977 & 0.8994 \\
		Llama~3.3   & 0.4675 & 0.1988 & 0.9782 & 0.3109 & 0.9045 \\
		Mistral Large & 1.1025 & 0.1431 & 1.6534 & 0.2722 & 0.9792 \\
		DeepSeek~V3 & 0.0087 & 0.2209 & 0.1146 & 0.1331 & 0.8653 \\
		\bottomrule
	\end{tabular}
\end{table*}

From the perspective of model fit, DeepSeek~V3 agents exhibit the best alignment with the theoretical models, yielding exceptionally low residual values (0.0087 and 0.1146). The simulated opinion updating process follows the DG dynamics almost perfectly. Although the DG model is a special case of the FJ model with $S = I$, we do not expect the residual sum in the former to be lower than in the latter. This has a simple explanation. From the outset, LLM agents under the FJ model receive a different prompt, constraining them to take their initial beliefs into consideration if they wish, whereas under the DG model the agents do not receive such information. Therefore, this information asymmetry leads to the observed outcome. Qwen2.5 agents achieve the next best fit after DeepSeek~V3 agents under both models, whereas Mistral Large agents demonstrate the weakest fit, consistent with their higher deviation from the model dynamics. This suggests that Mistral Large agents aggregate the information received from their peers in a different fashion. Group discussions with other types of agents (GPT-4o mini and Llama~3.3) achieve relatively fair alignment; however, some hidden rules in their opinion aggregation process may emerge.
		
Regarding the self-trust index, the results indicate variation in how strongly agents adhere to their prior beliefs. Higher values correspond to stronger opinion anchoring. Unlike the alignment in the sum of residuals across LLM agent types, which holds for both models, the estimate of the average self-trust index behaves differently depending on the model. Qwen2.5 has the highest average value of 0.3395, suggesting stronger persistence of initial opinions in the DG model, while dropping to 0.1977 in the FJ model. This suggests that additional information about possible anchoring to initial beliefs reduces the self-trust weight from about one third to one fifth, making agents more open to external influence in shaping their opinions. Interestingly, other agent types (GPT-4o mini, Llama~3.3, and Mistral Large) demonstrate the opposite pattern: self-trust weights increase when such additional information is provided, with agents relying more on their peers' opinions.
		
In the FJ model, we additionally focus on the average susceptibility to influence from peers. Across LLM agent types, the average agents' susceptibilities are consistently high (over 80\%), demonstrating their openness to group discussion and their willingness to take others' opinions into account. Notably, Mistral Large agents display the highest susceptibility (0.9792), suggesting that they rely heavily on peers' inputs and almost ignore their own initial beliefs, whereas GPT-4o mini agents show the lowest susceptibility (0.8226), reflecting a relatively more conservative updating process.

\begin{figure*}[t]
			\scriptsize
			\centering
			\begin{tikzpicture}
				\begin{groupplot}[
					group style={
						group size=2 by 5,
						horizontal sep=1.2cm,
						vertical sep=1.2cm,
						y descriptions at=edge left,
						group name=my plots1
					},
					xmin=1,
					xmax=20,
					ymin=0,
					ymax=1,
					xtick={1,5,10,15,20},
					ytick={0,0.25,...,1},
					width=7cm,
					height=4.5cm,
					legend columns=5,
					legend cell align=center,
					legend style={
						draw=none,
						font=\scriptsize,
						cells={anchor=center},
						at={(0,-0.2)},
						anchor=north
					},
					xlabel near ticks,
					ylabel near ticks
					]
					
					\nextgroupplot[title={DG~model, GPT-4o mini}, ylabel={Opinions}]
					\pgfplotstableread[col sep=&, row sep=\\]{
						1 & 2 & 3 & 4 & 5 & 6 & 7 & 8 & 9 & 10 & 11 & 12 & 13 & 14 & 15 & 16 & 17 & 18 & 19 & 20\\
						0.976593 & 0.832621 & 0.622122 & 0.464988 & 0.598028 & 0.761195 & 0.825395 & 0.642131 & 0.724047 & 0.702946 & 0.667636 & 0.795727 & 0.797733 & 0.749831 & 0.830695 & 0.767457 & 0.774502 & 0.794323 & 0.745139 & 0.769375 \\
						0.964542 & 0.472061 & 0.533916 & 0.712408 & 0.717132 & 0.695043 & 0.739689 & 0.774576 & 0.757217 & 0.745496 & 0.695661 & 0.758768 & 0.763471 & 0.804632 & 0.777993 & 0.793827 & 0.729592 & 0.79785 & 0.767337 & 0.748079 \\
						0.506483 & 0.674737 & 0.820601 & 0.646412 & 0.722668 & 0.656811 & 0.66295 & 0.67582 & 0.728571 & 0.718626 & 0.753524 & 0.777718 & 0.794343 & 0.634323 & 0.799184 & 0.735623 & 0.724224 & 0.702128 & 0.788757 & 0.713381 \\
						0.110872 & 0.344406 & 0.601615 & 0.730332 & 0.788304 & 0.740127 & 0.607412 & 0.729059 & 0.790112 & 0.862915 & 0.808561 & 0.753054 & 0.793894 & 0.824732 & 0.737434 & 0.753566 & 0.711655 & 0.7783 & 0.812814 & 0.7325 \\
						0.051001 & 0.589051 & 0.667247 & 0.63083 & 0.723312 & 0.687737 & 0.611827 & 0.766617 & 0.746408 & 0.688605 & 0.879129 & 0.689703 & 0.76599 & 0.780042 & 0.802376 & 0.668801 & 0.76736 & 0.740847 & 0.691546 & 0.681285 \\
						0.976593 & 0.813345 & 0.598185 & 0.578811 & 0.590308 & 0.595391 & 0.595705 & 0.595337 & 0.595209 & 0.595208 & 0.595219 & 0.595222 & 0.595222 & 0.595222 & 0.595221 & 0.595221 & 0.595222 & 0.595222 & 0.595222 & 0.595222 \\
						0.964542 & 0.486571 & 0.56837 & 0.594061 & 0.597672 & 0.595951 & 0.595202 & 0.59515 & 0.595204 & 0.595223 & 0.595224 & 0.595222 & 0.595221 & 0.595221 & 0.595221 & 0.595222 & 0.595222 & 0.595222 & 0.595222 & 0.595222 \\
						0.506483 & 0.661431 & 0.666847 & 0.603641 & 0.591548 & 0.59343 & 0.595073 & 0.595342 & 0.595267 & 0.595223 & 0.595218 & 0.59522 & 0.595222 & 0.595222 & 0.595222 & 0.595222 & 0.595222 & 0.595222 & 0.595222 & 0.595222 \\
						0.110872 & 0.363948 & 0.547233 & 0.606492 & 0.602227 & 0.596131 & 0.594836 & 0.595036 & 0.595206 & 0.595234 & 0.595226 & 0.595222 & 0.595221 & 0.595221 & 0.595222 & 0.595222 & 0.595222 & 0.595222 & 0.595222 & 0.595222 \\
						0.051001 & 0.609096 & 0.581072 & 0.596944 & 0.596294 & 0.595421 & 0.595173 & 0.595191 & 0.595218 & 0.595223 & 0.595222 & 0.595222 & 0.595221 & 0.595221 & 0.595222 & 0.595222 & 0.595222 & 0.595222 & 0.595222 & 0.595222 \\
					}\datatable;
					\pgfplotstabletranspose\datatable{\datatable};
					\addplot[mark=o, blue!60!black, thick, opacity=0.2, forget plot] table [x index=1, y index=2]{\datatable};
					\addplot[mark=o, red!90!black, thick, opacity=0.2, forget plot] table [x index=1, y index=3]{\datatable};
					\addplot[mark=o, orange!80!yellow, thick, opacity=0.2, forget plot] table [x index=1, y index=4]{\datatable};
					\addplot[mark=o, green!60!black, thick, opacity=0.2, forget plot] table [x index=1, y index=5]{\datatable};
					\addplot[mark=o, cyan!90!black, thick, opacity=0.2, forget plot] table [x index=1, y index=6]{\datatable};
					\addplot[mark=o, blue!60!black, thick] table [x index=1, y index=7]{\datatable};
					\addplot[mark=o, red!90!black, thick] table [x index=1, y index=8]{\datatable};
					\addplot[mark=o, orange!80!yellow, thick] table [x index=1, y index=9]{\datatable};
					\addplot[mark=o, green!60!black, thick] table [x index=1, y index=10]{\datatable};
					\addplot[mark=o, cyan!90!black, thick] table [x index=1, y index=11]{\datatable};
					
					\nextgroupplot[title={FJ~model, GPT-4o mini}]
					\pgfplotstableread[col sep=&, row sep=\\]{
						1 & 2 & 3 & 4 & 5 & 6 & 7 & 8 & 9 & 10 & 11 & 12 & 13 & 14 & 15 & 16 & 17 & 18 & 19 & 20\\
						0.980139 & 0.356776 & 0.426367 & 0.471784 & 0.533012 & 0.392018 & 0.467117 & 0.47929 & 0.40662 & 0.294864 & 0.360816 & 0.352242 & 0.534312 & 0.513853 & 0.564899 & 0.421932 & 0.434294 & 0.436304 & 0.362715 & 0.345195 \\
						0.977804 & 0.298349 & 0.54051 & 0.692853 & 0.577358 & 0.32232 & 0.40289 & 0.405443 & 0.512737 & 0.419638 & 0.409825 & 0.458432 & 0.314859 & 0.522192 & 0.569224 & 0.317421 & 0.390446 & 0.437799 & 0.419468 & 0.528538 \\
						0.470567 & 0.407289 & 0.601208 & 0.635809 & 0.239084 & 0.519982 & 0.557983 & 0.441291 & 0.457267 & 0.500696 & 0.334647 & 0.381167 & 0.443795 & 0.408797 & 0.519976 & 0.484322 & 0.490732 & 0.38916 & 0.484962 & 0.463684 \\
						0.098387 & 0.394275 & 0.585977 & 0.503648 & 0.344014 & 0.428894 & 0.411619 & 0.540388 & 0.539367 & 0.552646 & 0.342384 & 0.29108 & 0.332134 & 0.386515 & 0.331908 & 0.337559 & 0.466326 & 0.450844 & 0.351148 & 0.371856 \\
						0.045953 & 0.417511 & 0.51288 & 0.494308 & 0.486519 & 0.463238 & 0.3233 & 0.467156 & 0.344416 & 0.366775 & 0.280481 & 0.353579 & 0.400215 & 0.483625 & 0.474988 & 0.38724 & 0.422435 & 0.39761 & 0.412346 & 0.528012 \\
						0.980139 & 0.370695 & 0.407366 & 0.413543 & 0.416976 & 0.420308 & 0.423047 & 0.425339 & 0.427256 & 0.428859 & 0.430199 & 0.431319 & 0.432256 & 0.433039 & 0.433694 & 0.434242 & 0.4347 & 0.435082 & 0.435402 & 0.43567 \\
						0.977804 & 0.280671 & 0.433453 & 0.435725 & 0.436993 & 0.4398 & 0.442217 & 0.444391 & 0.446294 & 0.447936 & 0.449339 & 0.450531 & 0.451538 & 0.452386 & 0.4531 & 0.453699 & 0.454201 & 0.454621 & 0.454973 & 0.455268 \\
						0.470567 & 0.390006 & 0.458831 & 0.451841 & 0.453475 & 0.454425 & 0.455207 & 0.455866 & 0.456417 & 0.456877 & 0.457262 & 0.457584 & 0.457853 & 0.458078 & 0.458266 & 0.458423 & 0.458555 & 0.458665 & 0.458757 & 0.458834 \\
						0.098387 & 0.402442 & 0.381532 & 0.383162 & 0.38601 & 0.388927 & 0.391825 & 0.394506 & 0.396903 & 0.399 & 0.400808 & 0.402353 & 0.403665 & 0.404773 & 0.405707 & 0.406491 & 0.40715 & 0.407702 & 0.408165 & 0.408552 \\
						0.045953 & 0.42543 & 0.389313 & 0.39634 & 0.400441 & 0.403766 & 0.406577 & 0.408924 & 0.410887 & 0.412528 & 0.4139 & 0.415047 & 0.416006 & 0.416808 & 0.417479 & 0.41804 & 0.418508 & 0.4189 & 0.419228 & 0.419502 \\
					}\datatable;
					\pgfplotstabletranspose\datatable{\datatable};
					\addplot[mark=o, blue!60!black, thick, opacity=0.2, forget plot] table [x index=1, y index=2]{\datatable};
					\addplot[mark=o, red!90!black, thick, opacity=0.2, forget plot] table [x index=1, y index=3]{\datatable};
					\addplot[mark=o, orange!80!yellow, thick, opacity=0.2, forget plot] table [x index=1, y index=4]{\datatable};
					\addplot[mark=o, green!60!black, thick, opacity=0.2, forget plot] table [x index=1, y index=5]{\datatable};
					\addplot[mark=o, cyan!90!black, thick, opacity=0.2, forget plot] table [x index=1, y index=6]{\datatable};
					\addplot[mark=o, blue!60!black, thick] table [x index=1, y index=7]{\datatable};
					\addplot[mark=o, red!90!black, thick] table [x index=1, y index=8]{\datatable};
					\addplot[mark=o, orange!80!yellow, thick] table [x index=1, y index=9]{\datatable};
					\addplot[mark=o, green!60!black, thick] table [x index=1, y index=10]{\datatable};
					\addplot[mark=o, cyan!90!black, thick] table [x index=1, y index=11]{\datatable};

					\nextgroupplot[title={DG~model, Qwen2.5}, ylabel={Opinions}]
					\pgfplotstableread[col sep=&, row sep=\\]{
						1 & 2 & 3 & 4 & 5 & 6 & 7 & 8 & 9 & 10 & 11 & 12 & 13 & 14 & 15 & 16 & 17 & 18 & 19 & 20\\
						0.933106 & 0.516948 & 0.269485 & 0.276503 & 0.238593 & 0.165393 & 0.165393 & 0.116492 & 0.190864 & 0.214924 & 0.214924 & 0.209087 & 0.209087 & 0.209087 & 0.209087 & 0.265106 & 0.265106 & 0.223546 & 0.221386 & 0.198343 \\
						0.877179 & 0.366991 & 0.134419 & 0.19247 & 0.177954 & 0.165195 & 0.165195 & 0.187447 & 0.176165 & 0.214924 & 0.214924 & 0.209087 & 0.209087 & 0.203171 & 0.240764 & 0.25855 & 0.25855 & 0.221759 & 0.214011 & 0.216764 \\
						0.387907 & 0.186997 & 0.166772 & 0.190489 & 0.162863 & 0.185203 & 0.188154 & 0.181832 & 0.190979 & 0.214924 & 0.209087 & 0.179229 & 0.209087 & 0.209087 & 0.188612 & 0.207969 & 0.230359 & 0.202146 & 0.201437 & 0.212725 \\
						0.052616 & 0.084631 & 0.192732 & 0.191628 & 0.210464 & 0.161419 & 0.162989 & 0.20563 & 0.201428 & 0.201428 & 0.127153 & 0.209087 & 0.209087 & 0.209744 & 0.249351 & 0.184272 & 0.272067 & 0.239769 & 0.224128 & 0.212725 \\
						0.047125 & 0.180527 & 0.342637 & 0.181094 & 0.174416 & 0.145021 & 0.166852 & 0.176165 & 0.190979 & 0.214924 & 0.209087 & 0.209087 & 0.209087 & 0.105657 & 0.202346 & 0.252716 & 0.256131 & 0.272067 & 0.215217 & 0.209592 \\
						0.933106 & 0.507438 & 0.317235 & 0.237337 & 0.212637 & 0.207686 & 0.207833 & 0.208572 & 0.209014 & 0.209192 & 0.209245 & 0.209255 & 0.209255 & 0.209253 & 0.209252 & 0.209251 & 0.209251 & 0.209251 & 0.209251 & 0.209251 \\
						0.877179 & 0.356124 & 0.203678 & 0.18652 & 0.195794 & 0.203838 & 0.207639 & 0.208955 & 0.209278 & 0.209306 & 0.209282 & 0.209263 & 0.209255 & 0.209252 & 0.209251 & 0.209251 & 0.209251 & 0.209251 & 0.209251 & 0.209251 \\
						0.387907 & 0.20426 & 0.184914 & 0.195043 & 0.203608 & 0.207596 & 0.208959 & 0.209286 & 0.209311 & 0.209284 & 0.209264 & 0.209255 & 0.209252 & 0.209251 & 0.209251 & 0.209251 & 0.209251 & 0.209251 & 0.209251 & 0.209251 \\
						0.052616 & 0.130242 & 0.181032 & 0.201389 & 0.208 & 0.209496 & 0.209567 & 0.209416 & 0.209313 & 0.209268 & 0.209254 & 0.209251 & 0.209251 & 0.209251 & 0.209251 & 0.209251 & 0.209251 & 0.209251 & 0.209251 & 0.209251 \\
						0.047125 & 0.238825 & 0.244451 & 0.227854 & 0.21627 & 0.211194 & 0.209544 & 0.209179 & 0.209168 & 0.209209 & 0.209236 & 0.209247 & 0.209251 & 0.209251 & 0.209251 & 0.209251 & 0.209251 & 0.209251 & 0.209251 & 0.209251 \\
					}\datatable;
					\pgfplotstabletranspose\datatable{\datatable};
					\addplot[mark=o, blue!60!black, thick, opacity=0.2, forget plot] table [x index=1, y index=2]{\datatable};
					\addplot[mark=o, red!90!black, thick, opacity=0.2, forget plot] table [x index=1, y index=3]{\datatable};
					\addplot[mark=o, orange!80!yellow, thick, opacity=0.2, forget plot] table [x index=1, y index=4]{\datatable};
					\addplot[mark=o, green!60!black, thick, opacity=0.2, forget plot] table [x index=1, y index=5]{\datatable};
					\addplot[mark=o, cyan!90!black, thick, opacity=0.2, forget plot] table [x index=1, y index=6]{\datatable};
					\addplot[mark=o, blue!60!black, thick] table [x index=1, y index=7]{\datatable};
					\addplot[mark=o, red!90!black, thick] table [x index=1, y index=8]{\datatable};
					\addplot[mark=o, orange!80!yellow, thick] table [x index=1, y index=9]{\datatable};
					\addplot[mark=o, green!60!black, thick] table [x index=1, y index=10]{\datatable};
					\addplot[mark=o, cyan!90!black, thick] table [x index=1, y index=11]{\datatable};
					
					\nextgroupplot[title={FJ~model, Qwen2.5}]
					\pgfplotstableread[col sep=&, row sep=\\]{
						1 & 2 & 3 & 4 & 5 & 6 & 7 & 8 & 9 & 10 & 11 & 12 & 13 & 14 & 15 & 16 & 17 & 18 & 19 & 20\\
						0.936408 & 0.236719 & 0.465899 & 0.322371 & 0.267368 & 0.226892 & 0.200307 & 0.175702 & 0.181696 & 0.268713 & 0.262675 & 0.24629 & 0.27674 & 0.273173 & 0.282567 & 0.292322 & 0.20616 & 0.2784 & 0.274662 & 0.296626 \\
						0.770581 & 0.205646 & 0.264051 & 0.266075 & 0.359358 & 0.267116 & 0.200307 & 0.255598 & 0.181696 & 0.484719 & 0.343018 & 0.307733 & 0.240446 & 0.329745 & 0.297458 & 0.399555 & 0.379493 & 0.398575 & 0.216966 & 0.338154 \\
						0.425442 & 0.213796 & 0.165503 & 0.328538 & 0.331974 & 0.331974 & 0.182277 & 0.187256 & 0.181696 & 0.206841 & 0.269239 & 0.327179 & 0.242454 & 0.289724 & 0.297458 & 0.292322 & 0.2784 & 0.230945 & 0.197661 & 0.237802 \\
						0.061752 & 0.13627 & 0.219903 & 0.397681 & 0.247159 & 0.19076 & 0.159745 & 0.113171 & 0.181696 & 0.202167 & 0.238726 & 0.217429 & 0.25835 & 0.211523 & 0.27485 & 0.434335 & 0.220199 & 0.202004 & 0.243525 & 0.220599 \\
						0.063239 & 0.382299 & 0.317777 & 0.142338 & 0.204392 & 0.201083 & 0.20749 & 0.181696 & 0.182357 & 0.12469 & 0.24629 & 0.184242 & 0.205292 & 0.237918 & 0.2498 & 0.273173 & 0.292322 & 0.248237 & 0.246803 & 0.193831 \\
						0.936408 & 0.243101 & 0.35994 & 0.276361 & 0.266009 & 0.275128 & 0.273816 & 0.269805 & 0.267854 & 0.266578 & 0.265194 & 0.263911 & 0.262828 & 0.261887 & 0.261052 & 0.260318 & 0.259674 & 0.259109 & 0.258612 & 0.258175 \\
						0.770581 & 0.194051 & 0.306153 & 0.336365 & 0.325612 & 0.317381 & 0.316455 & 0.31525 & 0.313174 & 0.31143 & 0.310063 & 0.308845 & 0.307749 & 0.306789 & 0.305951 & 0.305213 & 0.304565 & 0.303995 & 0.303495 & 0.303055 \\
						0.425442 & 0.242173 & 0.245659 & 0.291498 & 0.274522 & 0.263732 & 0.263486 & 0.262525 & 0.260195 & 0.258296 & 0.256867 & 0.255583 & 0.254417 & 0.253398 & 0.252508 & 0.251726 & 0.251038 & 0.250434 & 0.249903 & 0.249437 \\
						0.061752 & 0.182183 & 0.277779 & 0.258087 & 0.23866 & 0.238671 & 0.238685 & 0.235922 & 0.233542 & 0.231901 & 0.230445 & 0.229092 & 0.227907 & 0.226879 & 0.225974 & 0.225177 & 0.224478 & 0.223863 & 0.223324 & 0.22285 \\
						0.063239 & 0.357143 & 0.225434 & 0.222597 & 0.237989 & 0.235571 & 0.230516 & 0.228732 & 0.227592 & 0.226152 & 0.224825 & 0.223731 & 0.222779 & 0.221931 & 0.221185 & 0.220531 & 0.219957 & 0.219453 & 0.219009 & 0.21862 \\
					}\datatable;
					\pgfplotstabletranspose\datatable{\datatable};
					\addplot[mark=o, blue!60!black, thick, opacity=0.2, forget plot] table [x index=1, y index=2]{\datatable};
					\addplot[mark=o, red!90!black, thick, opacity=0.2, forget plot] table [x index=1, y index=3]{\datatable};
					\addplot[mark=o, orange!80!yellow, thick, opacity=0.2, forget plot] table [x index=1, y index=4]{\datatable};
					\addplot[mark=o, green!60!black, thick, opacity=0.2, forget plot] table [x index=1, y index=5]{\datatable};
					\addplot[mark=o, cyan!90!black, thick, opacity=0.2, forget plot] table [x index=1, y index=6]{\datatable};
					\addplot[mark=o, blue!60!black, thick] table [x index=1, y index=7]{\datatable};
					\addplot[mark=o, red!90!black, thick] table [x index=1, y index=8]{\datatable};
					\addplot[mark=o, orange!80!yellow, thick] table [x index=1, y index=9]{\datatable};
					\addplot[mark=o, green!60!black, thick] table [x index=1, y index=10]{\datatable};
					\addplot[mark=o, cyan!90!black, thick] table [x index=1, y index=11]{\datatable};

					\nextgroupplot[title={DG~model, Llama~3.3}, ylabel={Opinions}]
					\pgfplotstableread[col sep=&, row sep=\\]{
						1 & 2 & 3 & 4 & 5 & 6 & 7 & 8 & 9 & 10 & 11 & 12 & 13 & 14 & 15 & 16 & 17 & 18 & 19 & 20\\
						0.953835 & 0.24026 & 0.284158 & 0.481184 & 0.519499 & 0.64607 & 0.610224 & 0.748558 & 0.677062 & 0.773281 & 0.728617 & 0.670244 & 0.668104 & 0.721014 & 0.746904 & 0.680093 & 0.684557 & 0.679426 & 0.787412 & 0.732402 \\
						0.958355 & 0.411136 & 0.618783 & 0.528072 & 0.511375 & 0.631571 & 0.776074 & 0.691892 & 0.640789 & 0.666964 & 0.748022 & 0.713443 & 0.597837 & 0.710401 & 0.754144 & 0.686546 & 0.763563 & 0.730226 & 0.786658 & 0.796838 \\
						0.267894 & 0.308324 & 0.523388 & 0.521143 & 0.566633 & 0.734806 & 0.657151 & 0.732374 & 0.646321 & 0.745287 & 0.799183 & 0.707245 & 0.712479 & 0.802025 & 0.674542 & 0.747943 & 0.708324 & 0.794759 & 0.752126 & 0.696034 \\
						0.041717 & 0.253295 & 0.376764 & 0.589071 & 0.608699 & 0.757919 & 0.717375 & 0.674196 & 0.644383 & 0.733351 & 0.674439 & 0.622992 & 0.755916 & 0.697274 & 0.714483 & 0.692142 & 0.710392 & 0.73991 & 0.825586 & 0.707118 \\
						0.052454 & 0.282285 & 0.28484 & 0.46222 & 0.464918 & 0.73859 & 0.623754 & 0.624734 & 0.781459 & 0.644762 & 0.747983 & 0.651309 & 0.704478 & 0.723259 & 0.805196 & 0.737696 & 0.718073 & 0.714513 & 0.706195 & 0.761574 \\
						0.953835 & 0.251479 & 0.320443 & 0.330064 & 0.333317 & 0.333922 & 0.334024 & 0.334044 & 0.334047 & 0.334048 & 0.334048 & 0.334048 & 0.334048 & 0.334048 & 0.334048 & 0.334048 & 0.334048 & 0.334048 & 0.334048 & 0.334048 \\
						0.958355 & 0.426719 & 0.350308 & 0.336447 & 0.334471 & 0.33411 & 0.334059 & 0.334049 & 0.334048 & 0.334048 & 0.334048 & 0.334048 & 0.334048 & 0.334048 & 0.334048 & 0.334048 & 0.334048 & 0.334048 & 0.334048 & 0.334048 \\
						0.267894 & 0.336031 & 0.332255 & 0.334112 & 0.333999 & 0.33405 & 0.334047 & 0.334048 & 0.334048 & 0.334048 & 0.334048 & 0.334048 & 0.334048 & 0.334048 & 0.334048 & 0.334048 & 0.334048 & 0.334048 & 0.334048 & 0.334048 \\
						0.041717 & 0.279344 & 0.326501 & 0.33262 & 0.333853 & 0.334011 & 0.334043 & 0.334047 & 0.334048 & 0.334048 & 0.334048 & 0.334048 & 0.334048 & 0.334048 & 0.334048 & 0.334048 & 0.334048 & 0.334048 & 0.334048 & 0.334048 \\
						0.052454 & 0.298059 & 0.304284 & 0.328538 & 0.332997 & 0.333846 & 0.334017 & 0.334042 & 0.334047 & 0.334048 & 0.334048 & 0.334048 & 0.334048 & 0.334048 & 0.334048 & 0.334048 & 0.334048 & 0.334048 & 0.334048 & 0.334048 \\
					}\datatable;
					\pgfplotstabletranspose\datatable{\datatable};
					\addplot[mark=o, blue!60!black, thick, opacity=0.2, forget plot] table [x index=1, y index=2]{\datatable};
					\addplot[mark=o, red!90!black, thick, opacity=0.2, forget plot] table [x index=1, y index=3]{\datatable};
					\addplot[mark=o, orange!80!yellow, thick, opacity=0.2, forget plot] table [x index=1, y index=4]{\datatable};
					\addplot[mark=o, green!60!black, thick, opacity=0.2, forget plot] table [x index=1, y index=5]{\datatable};
					\addplot[mark=o, cyan!90!black, thick, opacity=0.2, forget plot] table [x index=1, y index=6]{\datatable};
					\addplot[mark=o, blue!60!black, thick] table [x index=1, y index=7]{\datatable};
					\addplot[mark=o, red!90!black, thick] table [x index=1, y index=8]{\datatable};
					\addplot[mark=o, orange!80!yellow, thick] table [x index=1, y index=9]{\datatable};
					\addplot[mark=o, green!60!black, thick] table [x index=1, y index=10]{\datatable};
					\addplot[mark=o, cyan!90!black, thick] table [x index=1, y index=11]{\datatable};
					
					\nextgroupplot[title={FJ~model, Llama~3.3}]
					\pgfplotstableread[col sep=&, row sep=\\]{
						1 & 2 & 3 & 4 & 5 & 6 & 7 & 8 & 9 & 10 & 11 & 12 & 13 & 14 & 15 & 16 & 17 & 18 & 19 & 20\\
						0.971498 & 0.489515 & 0.499792 & 0.640036 & 0.489526 & 0.377736 & 0.65459 & 0.565285 & 0.661105 & 0.643447 & 0.594298 & 0.546147 & 0.641407 & 0.553721 & 0.550282 & 0.589111 & 0.588687 & 0.537935 & 0.504031 & 0.558939 \\
						0.950211 & 0.620652 & 0.515602 & 0.502407 & 0.489157 & 0.285327 & 0.358361 & 0.535673 & 0.265854 & 0.550197 & 0.437162 & 0.648329 & 0.586102 & 0.523007 & 0.383379 & 0.407165 & 0.519538 & 0.478244 & 0.483879 & 0.505884 \\
						0.245774 & 0.50127 & 0.437657 & 0.52837 & 0.242704 & 0.150987 & 0.554638 & 0.42096 & 0.385783 & 0.499259 & 0.558509 & 0.524309 & 0.535919 & 0.493924 & 0.49965 & 0.474125 & 0.557404 & 0.541014 & 0.570308 & 0.533339 \\
						0.082851 & 0.214972 & 0.535308 & 0.613446 & 0.590779 & 0.390531 & 0.550325 & 0.597365 & 0.247684 & 0.67359 & 0.528261 & 0.549737 & 0.488994 & 0.449353 & 0.456325 & 0.4049 & 0.401711 & 0.447456 & 0.50949 & 0.572652 \\
						0.038845 & 0.308748 & 0.27025 & 0.222012 & 0.222012 & 0.151454 & 0.319315 & 0.629442 & 0.535591 & 0.483257 & 0.577712 & 0.477008 & 0.366856 & 0.517295 & 0.514441 & 0.372588 & 0.313276 & 0.418908 & 0.490793 & 0.501 \\
						0.971498 & 0.498674 & 0.512778 & 0.541997 & 0.551822 & 0.559841 & 0.567353 & 0.57398 & 0.579754 & 0.584769 & 0.589118 & 0.592884 & 0.596146 & 0.59897 & 0.601416 & 0.603533 & 0.605366 & 0.606953 & 0.608327 & 0.609516 \\
						0.950211 & 0.574342 & 0.503313 & 0.465249 & 0.466249 & 0.475078 & 0.483956 & 0.492002 & 0.499098 & 0.505282 & 0.510648 & 0.515299 & 0.519327 & 0.522814 & 0.525834 & 0.528449 & 0.530712 & 0.532672 & 0.534369 & 0.535838 \\
						0.245774 & 0.4996 & 0.431476 & 0.424674 & 0.434717 & 0.443976 & 0.452052 & 0.45914 & 0.465304 & 0.470646 & 0.475274 & 0.479281 & 0.48275 & 0.485754 & 0.488355 & 0.490606 & 0.492556 & 0.494243 & 0.495704 & 0.49697 \\
						0.082851 & 0.339253 & 0.442844 & 0.441508 & 0.443583 & 0.451057 & 0.459317 & 0.467117 & 0.474126 & 0.480287 & 0.485654 & 0.490312 & 0.494348 & 0.497844 & 0.500872 & 0.503493 & 0.505762 & 0.507727 & 0.509428 & 0.510901 \\
						0.038845 & 0.313129 & 0.355224 & 0.37151 & 0.386571 & 0.397939 & 0.406889 & 0.414353 & 0.420717 & 0.426192 & 0.430919 & 0.435007 & 0.438545 & 0.441608 & 0.444259 & 0.446555 & 0.448542 & 0.450262 & 0.451752 & 0.453042 \\
					}\datatable;
					\pgfplotstabletranspose\datatable{\datatable};
					\addplot[mark=o, blue!60!black, thick, opacity=0.2, forget plot] table [x index=1, y index=2]{\datatable};
					\addplot[mark=o, red!90!black, thick, opacity=0.2, forget plot] table [x index=1, y index=3]{\datatable};
					\addplot[mark=o, orange!80!yellow, thick, opacity=0.2, forget plot] table [x index=1, y index=4]{\datatable};
					\addplot[mark=o, green!60!black, thick, opacity=0.2, forget plot] table [x index=1, y index=5]{\datatable};
					\addplot[mark=o, cyan!90!black, thick, opacity=0.2, forget plot] table [x index=1, y index=6]{\datatable};
					\addplot[mark=o, blue!60!black, thick] table [x index=1, y index=7]{\datatable};
					\addplot[mark=o, red!90!black, thick] table [x index=1, y index=8]{\datatable};
					\addplot[mark=o, orange!80!yellow, thick] table [x index=1, y index=9]{\datatable};
					\addplot[mark=o, green!60!black, thick] table [x index=1, y index=10]{\datatable};
					\addplot[mark=o, cyan!90!black, thick] table [x index=1, y index=11]{\datatable};

					\nextgroupplot[title={DG~model, Mistral Large}, ylabel={Opinions}]
					\pgfplotstableread[col sep=&, row sep=\\]{
						1 & 2 & 3 & 4 & 5 & 6 & 7 & 8 & 9 & 10 & 11 & 12 & 13 & 14 & 15 & 16 & 17 & 18 & 19 & 20\\
						0.965606 & 0.459423 & 0.392713 & 0.52995 & 0.457663 & 0.669614 & 0.604547 & 0.406639 & 0.473568 & 0.448114 & 0.476987 & 0.320652 & 0.292913 & 0.487953 & 0.551231 & 0.392299 & 0.510243 & 0.5081 & 0.537199 & 0.415285 \\
						0.960496 & 0.363695 & 0.36396 & 0.572647 & 0.551519 & 0.351741 & 0.553226 & 0.552849 & 0.499924 & 0.527565 & 0.511931 & 0.290322 & 0.380849 & 0.423447 & 0.545634 & 0.414027 & 0.574396 & 0.375003 & 0.491482 & 0.413933 \\
						0.157702 & 0.537174 & 0.266606 & 0.321512 & 0.379471 & 0.583939 & 0.679179 & 0.62138 & 0.557255 & 0.495421 & 0.488166 & 0.287143 & 0.605418 & 0.489628 & 0.582561 & 0.597368 & 0.574902 & 0.318482 & 0.493514 & 0.395374 \\
						0.098445 & 0.603604 & 0.446503 & 0.593267 & 0.410263 & 0.746034 & 0.505974 & 0.361269 & 0.399135 & 0.588266 & 0.317532 & 0.509822 & 0.339197 & 0.354183 & 0.439835 & 0.455194 & 0.511638 & 0.55211 & 0.555641 & 0.414206 \\
						0.042677 & 0.748506 & 0.502675 & 0.581989 & 0.572186 & 0.604558 & 0.652552 & 0.330242 & 0.409261 & 0.495201 & 0.516456 & 0.274833 & 0.32567 & 0.326016 & 0.528807 & 0.625237 & 0.549797 & 0.443206 & 0.535137 & 0.496932 \\
						0.965606 & 0.476148 & 0.504656 & 0.520365 & 0.515417 & 0.515585 & 0.515832 & 0.515739 & 0.515754 & 0.515752 & 0.515752 & 0.515752 & 0.515752 & 0.515752 & 0.515752 & 0.515752 & 0.515752 & 0.515752 & 0.515752 & 0.515752 \\
						0.960496 & 0.373049 & 0.527602 & 0.518622 & 0.514528 & 0.515748 & 0.515828 & 0.515718 & 0.515759 & 0.515751 & 0.515753 & 0.515752 & 0.515752 & 0.515752 & 0.515752 & 0.515752 & 0.515752 & 0.515752 & 0.515752 & 0.515752 \\
						0.157702 & 0.520336 & 0.511216 & 0.510287 & 0.515393 & 0.515557 & 0.515649 & 0.515738 & 0.515746 & 0.51575 & 0.515752 & 0.515752 & 0.515752 & 0.515752 & 0.515752 & 0.515752 & 0.515752 & 0.515752 & 0.515752 & 0.515752 \\
						0.098445 & 0.622522 & 0.516803 & 0.513673 & 0.516616 & 0.515992 & 0.515678 & 0.515792 & 0.515748 & 0.515755 & 0.515752 & 0.515753 & 0.515752 & 0.515752 & 0.515752 & 0.515752 & 0.515752 & 0.515752 & 0.515752 & 0.515752 \\
						0.042677 & 0.710767 & 0.50146 & 0.512121 & 0.518108 & 0.515617 & 0.515732 & 0.515786 & 0.515751 & 0.515753 & 0.515753 & 0.515752 & 0.515752 & 0.515752 & 0.515752 & 0.515752 & 0.515752 & 0.515752 & 0.515752 & 0.515752 \\
					}\datatable;
					\pgfplotstabletranspose\datatable{\datatable};
					\addplot[mark=o, blue!60!black, thick, opacity=0.2, forget plot] table [x index=1, y index=2]{\datatable};
					\addplot[mark=o, red!90!black, thick, opacity=0.2, forget plot] table [x index=1, y index=3]{\datatable};
					\addplot[mark=o, orange!80!yellow, thick, opacity=0.2, forget plot] table [x index=1, y index=4]{\datatable};
					\addplot[mark=o, green!60!black, thick, opacity=0.2, forget plot] table [x index=1, y index=5]{\datatable};
					\addplot[mark=o, cyan!90!black, thick, opacity=0.2, forget plot] table [x index=1, y index=6]{\datatable};
					\addplot[mark=o, blue!60!black, thick] table [x index=1, y index=7]{\datatable};
					\addplot[mark=o, red!90!black, thick] table [x index=1, y index=8]{\datatable};
					\addplot[mark=o, orange!80!yellow, thick] table [x index=1, y index=9]{\datatable};
					\addplot[mark=o, green!60!black, thick] table [x index=1, y index=10]{\datatable};
					\addplot[mark=o, cyan!90!black, thick] table [x index=1, y index=11]{\datatable};
					
					\nextgroupplot[title={FJ~model, Mistral Large}]
					\pgfplotstableread[col sep=&, row sep=\\]{
						1 & 2 & 3 & 4 & 5 & 6 & 7 & 8 & 9 & 10 & 11 & 12 & 13 & 14 & 15 & 16 & 17 & 18 & 19 & 20\\
						0.982942 & 0.392153 & 0.441644 & 0.428943 & 0.388177 & 0.335312 & 0.3283 & 0.32132 & 0.395433 & 0.400852 & 0.45512 & 0.606436 & 0.392899 & 0.316834 & 0.697936 & 0.706961 & 0.697936 & 0.857302 & 0.442898 & 0.341005 \\
						0.927808 & 0.236019 & 0.68428 & 0.492588 & 0.556098 & 0.370295 & 0.370295 & 0.383746 & 0.368992 & 0.567505 & 0.343356 & 0.348508 & 0.446022 & 0.617513 & 0.617513 & 0.617513 & 0.382185 & 0.382185 & 0.524948 & 0.261423 \\
						0.220514 & 0.225611 & 0.301207 & 0.71036 & 0.543945 & 0.471662 & 0.620704 & 0.326188 & 0.39271 & 0.463138 & 0.3116 & 0.374684 & 0.627778 & 0.720165 & 0.570767 & 0.617552 & 0.842395 & 0.570454 & 0.383567 & 0.383567 \\
						0.101666 & 0.189506 & 0.340584 & 0.351479 & 0.627244 & 0.631787 & 0.631787 & 0.365443 & 0.425214 & 0.222919 & 0.288819 & 0.609265 & 0.592604 & 0.373369 & 0.583529 & 0.741898 & 0.798563 & 0.654955 & 0.763559 & 0.538602 \\
						0.038147 & 0.449279 & 0.672268 & 0.647524 & 0.404341 & 0.376104 & 0.350341 & 0.327329 & 0.261066 & 0.400768 & 0.209072 & 0.492114 & 0.294341 & 0.59523 & 0.637568 & 0.755762 & 0.570315 & 0.516317 & 0.225271 & 0.270215 \\
						0.982942 & 0.428431 & 0.378029 & 0.361513 & 0.359599 & 0.355928 & 0.35233 & 0.348627 & 0.345002 & 0.341487 & 0.338101 & 0.334849 & 0.331729 & 0.328738 & 0.325871 & 0.323124 & 0.320491 & 0.317968 & 0.31555 & 0.313233 \\
						0.927808 & 0.293377 & 0.399715 & 0.378215 & 0.377709 & 0.372896 & 0.369186 & 0.365496 & 0.362007 & 0.358664 & 0.355465 & 0.352401 & 0.349465 & 0.346651 & 0.343956 & 0.341373 & 0.338898 & 0.336526 & 0.334253 & 0.332076 \\
						0.220514 & 0.250471 & 0.345469 & 0.357555 & 0.356712 & 0.353789 & 0.350303 & 0.346694 & 0.343116 & 0.339636 & 0.336277 & 0.333049 & 0.32995 & 0.326979 & 0.324131 & 0.321402 & 0.318786 & 0.31628 & 0.313878 & 0.311577 \\
						0.101666 & 0.250246 & 0.343048 & 0.351193 & 0.353431 & 0.351402 & 0.348387 & 0.344909 & 0.34134 & 0.337811 & 0.334381 & 0.331073 & 0.327894 & 0.324843 & 0.321918 & 0.319115 & 0.316427 & 0.313852 & 0.311385 & 0.30902 \\
						0.038147 & 0.44118 & 0.348243 & 0.350564 & 0.342647 & 0.33871 & 0.334706 & 0.331108 & 0.327697 & 0.324458 & 0.321365 & 0.318406 & 0.315574 & 0.31286 & 0.31026 & 0.307769 & 0.305382 & 0.303095 & 0.300903 & 0.298803 \\
					}\datatable;
					\pgfplotstabletranspose\datatable{\datatable};
					\addplot[mark=o, blue!60!black, thick, opacity=0.2, forget plot] table [x index=1, y index=2]{\datatable};
					\addplot[mark=o, red!90!black, thick, opacity=0.2, forget plot] table [x index=1, y index=3]{\datatable};
					\addplot[mark=o, orange!80!yellow, thick, opacity=0.2, forget plot] table [x index=1, y index=4]{\datatable};
					\addplot[mark=o, green!60!black, thick, opacity=0.2, forget plot] table [x index=1, y index=5]{\datatable};
					\addplot[mark=o, cyan!90!black, thick, opacity=0.2, forget plot] table [x index=1, y index=6]{\datatable};
					\addplot[mark=o, blue!60!black, thick] table [x index=1, y index=7]{\datatable};
					\addplot[mark=o, red!90!black, thick] table [x index=1, y index=8]{\datatable};
					\addplot[mark=o, orange!80!yellow, thick] table [x index=1, y index=9]{\datatable};
					\addplot[mark=o, green!60!black, thick] table [x index=1, y index=10]{\datatable};
					\addplot[mark=o, cyan!90!black, thick] table [x index=1, y index=11]{\datatable};

					\nextgroupplot[title={DG~model, DeepSeek~V3}, ylabel={Opinions}]
					\pgfplotstableread[col sep=&, row sep=\\]{
						1 & 2 & 3 & 4 & 5 & 6 & 7 & 8 & 9 & 10 & 11 & 12 & 13 & 14 & 15 & 16 & 17 & 18 & 19 & 20\\
						0.977667 & 0.20859 & 0.102306 & 0.095247 & 0.069807 & 0.05025 & 0.052871 & 0.069111 & 0.059505 & 0.061155 & 0.064223 & 0.064223 & 0.064223 & 0.070297 & 0.056701 & 0.049305 & 0.052892 & 0.048664 & 0.053048 & 0.045804 \\
						0.97001 & 0.324456 & 0.125711 & 0.054608 & 0.069989 & 0.059956 & 0.058002 & 0.069396 & 0.065803 & 0.066683 & 0.064223 & 0.067208 & 0.051137 & 0.071538 & 0.051302 & 0.055404 & 0.05714 & 0.053048 & 0.051152 & 0.07124 \\
						0.261009 & 0.13377 & 0.152933 & 0.082198 & 0.074172 & 0.059924 & 0.085451 & 0.055912 & 0.063282 & 0.06983 & 0.061775 & 0.064223 & 0.064223 & 0.060128 & 0.054322 & 0.048894 & 0.050461 & 0.049139 & 0.049139 & 0.063326 \\
						0.02891 & 0.049536 & 0.055112 & 0.061909 & 0.05568 & 0.051073 & 0.087672 & 0.054072 & 0.067444 & 0.064585 & 0.064585 & 0.064585 & 0.065082 & 0.057207 & 0.058157 & 0.051586 & 0.052892 & 0.051361 & 0.049139 & 0.048867 \\
						0.040847 & 0.1105 & 0.054114 & 0.062416 & 0.070506 & 0.060672 & 0.062533 & 0.079328 & 0.05934 & 0.077108 & 0.077108 & 0.077108 & 0.078393 & 0.048265 & 0.058235 & 0.050017 & 0.054694 & 0.049139 & 0.048126 & 0.073455 \\
						0.977667 & 0.208818 & 0.104427 & 0.0899759 & 0.0738057 & 0.0701791 & 0.068837 & 0.0683983 & 0.0682509 & 0.0682023 & 0.0681861 & 0.0681808 & 0.068179 & 0.0681784 & 0.0681782 & 0.0681781 & 0.0681781 & 0.0681781 & 0.0681781 & 0.0681781 \\
						0.97001 & 0.32416 & 0.12145 & 0.0779915 & 0.074176 & 0.0698222 & 0.0687473 & 0.0683659 & 0.0682409 & 0.0681989 & 0.068185 & 0.0681804 & 0.0681788 & 0.0681783 & 0.0681782 & 0.0681781 & 0.0681781 & 0.0681781 & 0.0681781 & 0.0681781 \\
						0.261009 & 0.140882 & 0.128322 & 0.0804657 & 0.0730437 & 0.0697501 & 0.0687057 & 0.0683519 & 0.0682362 & 0.0681973 & 0.0681845 & 0.0681802 & 0.0681788 & 0.0681783 & 0.0681782 & 0.0681781 & 0.0681781 & 0.0681781 & 0.0681781 & 0.0681781 \\
						0.02891 & 0.048516 & 0.0666208 & 0.0663039 & 0.0677451 & 0.0680169 & 0.0681261 & 0.0681605 & 0.0681723 & 0.0681761 & 0.0681774 & 0.0681779 & 0.068178 & 0.068178 & 0.0681781 & 0.0681781 & 0.0681781 & 0.0681781 & 0.0681781 & 0.0681781 \\
						0.040847 & 0.109525 & 0.0627162 & 0.0703986 & 0.0684059 & 0.0683013 & 0.0682137 & 0.0681909 & 0.0681822 & 0.0681795 & 0.0681785 & 0.0681782 & 0.0681781 & 0.0681781 & 0.0681781 & 0.0681781 & 0.0681781 & 0.0681781 & 0.0681781 & 0.0681781 \\
					}\datatable;
					\pgfplotstabletranspose\datatable{\datatable};
					\addplot[mark=o, blue!60!black, thick, opacity=0.2, forget plot] table [x index=1, y index=2]{\datatable};
					\addplot[mark=o, red!90!black, thick, opacity=0.2, forget plot] table [x index=1, y index=3]{\datatable};
					\addplot[mark=o, orange!80!yellow, thick, opacity=0.2, forget plot] table [x index=1, y index=4]{\datatable};
					\addplot[mark=o, green!60!black, thick, opacity=0.2, forget plot] table [x index=1, y index=5]{\datatable};
					\addplot[mark=o, cyan!90!black, thick, opacity=0.2, forget plot] table [x index=1, y index=6]{\datatable};
					\addplot[mark=o, blue!60!black, thick] table [x index=1, y index=7]{\datatable};
					\addplot[mark=o, red!90!black, thick] table [x index=1, y index=8]{\datatable};
					\addplot[mark=o, orange!80!yellow, thick] table [x index=1, y index=9]{\datatable};
					\addplot[mark=o, green!60!black, thick] table [x index=1, y index=10]{\datatable};
					\addplot[mark=o, cyan!90!black, thick] table [x index=1, y index=11]{\datatable};
					
					\nextgroupplot[title={FJ~model, DeepSeek~V3}]
					\pgfplotstableread[col sep=&, row sep=\\]{
						1 & 2 & 3 & 4 & 5 & 6 & 7 & 8 & 9 & 10 & 11 & 12 & 13 & 14 & 15 & 16 & 17 & 18 & 19 & 20\\
						0.979654 & 0.19821 & 0.13159 & 0.079232 & 0.070448 & 0.062251 & 0.11803 & 0.150897 & 0.046399 & 0.048521 & 0.118237 & 0.049631 & 0.038539 & 0.057762 & 0.031385 & 0.032051 & 0.046383 & 0.039126 & 0.049356 & 0.037911 \\
						0.957896 & 0.124634 & 0.081024 & 0.094965 & 0.141771 & 0.101545 & 0.069691 & 0.141532 & 0.154182 & 0.049991 & 0.057634 & 0.037376 & 0.044114 & 0.034726 & 0.037016 & 0.037528 & 0.034284 & 0.033139 & 0.064887 & 0.041751 \\
						0.489264 & 0.135673 & 0.278271 & 0.065132 & 0.22038 & 0.048956 & 0.187898 & 0.091834 & 0.065039 & 0.102901 & 0.040225 & 0.041657 & 0.049916 & 0.037195 & 0.0349 & 0.038748 & 0.057194 & 0.03558 & 0.035006 & 0.038631 \\
						0.034928 & 0.208467 & 0.100665 & 0.089566 & 0.090706 & 0.048791 & 0.086269 & 0.059425 & 0.045047 & 0.053645 & 0.051898 & 0.034402 & 0.039438 & 0.035169 & 0.036496 & 0.037762 & 0.040868 & 0.038282 & 0.039503 & 0.040474 \\
						0.032503 & 0.071863 & 0.104133 & 0.298714 & 0.090218 & 0.060686 & 0.052629 & 0.072113 & 0.051377 & 0.037762 & 0.039663 & 0.039885 & 0.039183 & 0.042263 & 0.037205 & 0.037767 & 0.039761 & 0.041934 & 0.040522 & 0.039177 \\
						0.979654 & 0.199206 & 0.189933 & 0.12223 & 0.103133 & 0.0863001 & 0.0740015 & 0.0651427 & 0.0586905 & 0.0540099 & 0.0506106 & 0.0481424 & 0.0463503 & 0.045049 & 0.0441042 & 0.0434181 & 0.04292 & 0.0425583 & 0.0422957 & 0.042105 \\
						0.957896 & 0.135782 & 0.15959 & 0.138565 & 0.112407 & 0.0933714 & 0.0797994 & 0.0699037 & 0.0627254 & 0.0575121 & 0.0537268 & 0.0509783 & 0.0489827 & 0.0475336 & 0.0464814 & 0.0457175 & 0.0451628 & 0.04476 & 0.0444675 & 0.0442552 \\
						0.489264 & 0.132315 & 0.188764 & 0.134938 & 0.116846 & 0.0994209 & 0.0874478 & 0.0786753 & 0.0723102 & 0.0676894 & 0.0643338 & 0.0618975 & 0.0601284 & 0.0588439 & 0.0579112 & 0.057234 & 0.0567422 & 0.0563852 & 0.0561259 & 0.0559377 \\
						0.034928 & 0.206931 & 0.106236 & 0.091461 & 0.0766268 & 0.0661606 & 0.0585489 & 0.0530097 & 0.048992 & 0.0460738 & 0.0439551 & 0.0424167 & 0.0412996 & 0.0404885 & 0.0398996 & 0.039472 & 0.0391615 & 0.038936 & 0.0387723 & 0.0386535 \\
						0.032503 & 0.125629 & 0.145678 & 0.118635 & 0.0959318 & 0.0813078 & 0.070339 & 0.0624338 & 0.056685 & 0.0525117 & 0.0494815 & 0.0472812 & 0.0456836 & 0.0445235 & 0.0436812 & 0.0430697 & 0.0426256 & 0.0423031 & 0.042069 & 0.041899 \\
					}\datatable;
					\pgfplotstabletranspose\datatable{\datatable};
					\addplot[mark=o, blue!60!black, thick, opacity=0.2] table [x index=1, y index=2]{\datatable};\addlegendentry{$x_1(t)\quad$}
					\addplot[mark=o, red!90!black, thick, opacity=0.2] table [x index=1, y index=3]{\datatable};\addlegendentry{$x_2(t)\quad$}
					\addplot[mark=o, orange!80!yellow, thick, opacity=0.2] table [x index=1, y index=4]{\datatable};\addlegendentry{$x_3(t)\quad$}
					\addplot[mark=o, green!60!black, thick, opacity=0.2] table [x index=1, y index=5]{\datatable};\addlegendentry{$x_4(t)\quad$}
					\addplot[mark=o, cyan!90!black, thick, opacity=0.2] table [x index=1, y index=6]{\datatable};\addlegendentry{$x_5(t)$}
					\addplot[mark=o, blue!60!black, thick] table [x index=1, y index=7]{\datatable};\addlegendentry{$\widehat{x}_1(t)\quad$}
					\addplot[mark=o, red!90!black, thick] table [x index=1, y index=8]{\datatable};\addlegendentry{$\widehat{x}_2(t)\quad$}
					\addplot[mark=o, orange!80!yellow, thick] table [x index=1, y index=9]{\datatable};\addlegendentry{$\widehat{x}_3(t)\quad$}
					\addplot[mark=o, green!60!black, thick] table [x index=1, y index=10]{\datatable};\addlegendentry{$\widehat{x}_4(t)\quad$}
					\addplot[mark=o, cyan!90!black, thick] table [x index=1, y index=11]{\datatable};\addlegendentry{$\widehat{x}_5(t)$}
				\end{groupplot}
			\end{tikzpicture}
	\caption{Components of the original opinion (sentiment) vector $x(t)$ and fitted opinion vector $\widehat{x}(t)$ for homogeneous populations of LLM agents under different opinion dynamics models. The horizontal axis represents the round number.}
	\label{fig:homogeneous}
\end{figure*}
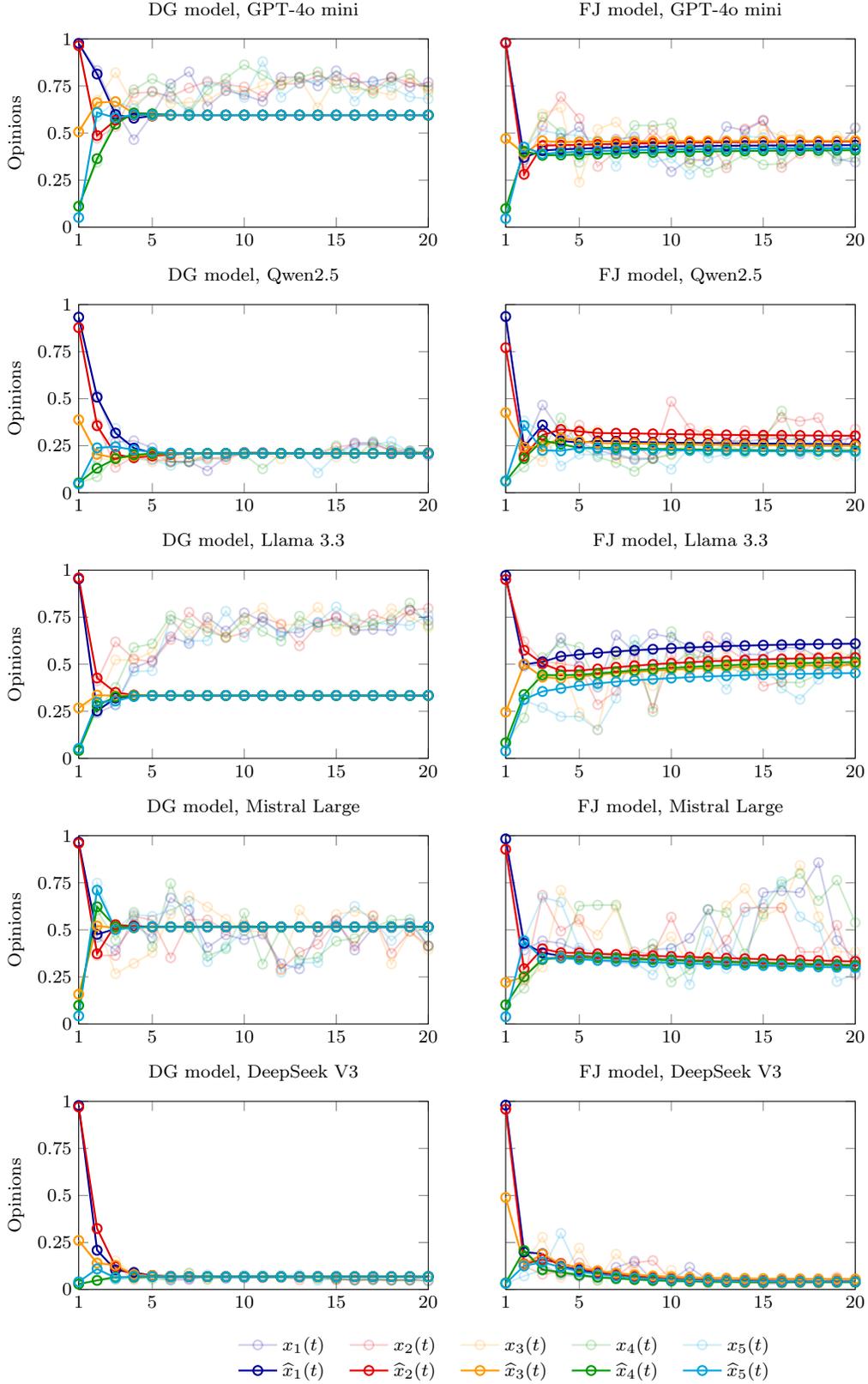

Overall, the DeGroot results reinforce the trends observed under the FJ framework: DeepSeek~V3 agents consistently produce opinion dynamics that closely adhere to classical models, whereas Mistral Large agents tend to diverge, likely due to their higher susceptibility to peer influence or other underlying mechanisms. Meanwhile, Qwen2.5 agents stand out for their relatively stronger self anchoring behavior, suggesting systematic differences in how various LLMs internalize and propagate opinions during deliberation. Fig.~\ref{fig:homogeneous} visualizes the fitted opinion trajectories, demonstrating that opinion formation by LLM agents is largely consistent with the FJ model (except for the Mistral Large case, which exhibits the poorest fit and likely reflects a different underlying opinion formation mechanism). Notably, the opinion dynamics of the Qwen2.5 and DeepSeek~V3 agents are also consistent with the DG model.

\subsection{Heterogeneous LLM agents}\label{sec4-2}

In the heterogeneous case, we again simulate a 20 round group discussion among five LLM agents, each having a distinct type and distinct initial stance parameters. Specifically, the simulation begins with the following stance configuration: DeepSeek~V3 is initialized as \emph{strongly positive}, GPT-4o mini as \emph{positive}, Qwen2.5 as \emph{neutral}, Mistral Large as \emph{negative}, and Llama~3.3 as \emph{strongly negative}.

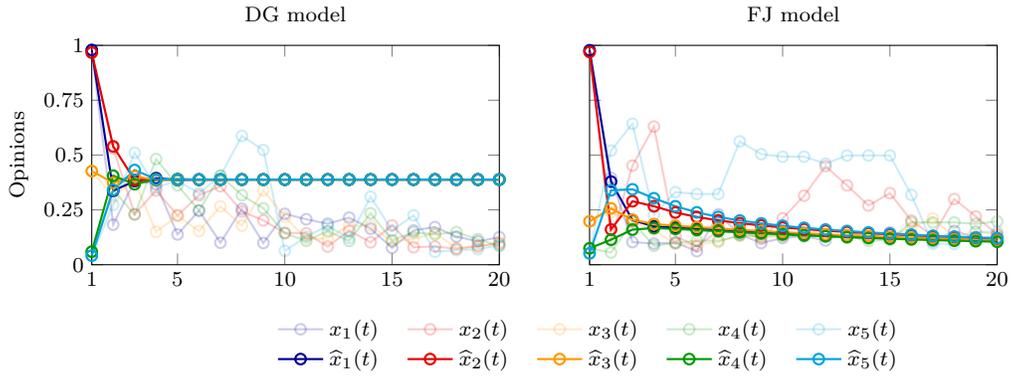
\begin{figure*}[t]
			\scriptsize
			\centering
			\begin{tikzpicture}
				\begin{groupplot}[
					group style={
						group size=2 by 5,
						horizontal sep=1.2cm,
						vertical sep=1.2cm,
						y descriptions at=edge left,
						group name=my plots2
					},
					xmin=1,
					xmax=20,
					ymin=0,
					ymax=1,
					xtick={1,5,10,15,20},
					ytick={0,0.25,...,1},
					width=7cm,
					height=4.5cm,
					legend columns=5,
					legend cell align=center,
					legend style={
						draw=none,
						font=\scriptsize,
						cells={anchor=center},
						at={(0,-0.2)},
						anchor=north
					},
					xlabel near ticks,
					ylabel near ticks
					]
					
					\nextgroupplot[title={DG~model}, ylabel={Opinions}]
					\pgfplotstableread[col sep=&, row sep=\\]{
						1 & 2 & 3 & 4 & 5 & 6 & 7 & 8 & 9 & 10 & 11 & 12 & 13 & 14 & 15 & 16 & 17 & 18 & 19 & 20\\
						0.978868 & 0.182642 & 0.384071 & 0.366558 & 0.138049 & 0.246562 & 0.100415 & 0.254914 & 0.098878 & 0.232076 & 0.207233 & 0.186789 & 0.214647 & 0.16443 & 0.076668 & 0.156157 & 0.171572 & 0.12822 & 0.106754 & 0.125504 \\
						0.967102 & 0.529602 & 0.230586 & 0.338536 & 0.225054 & 0.316719 & 0.35758 & 0.242787 & 0.202072 & 0.145493 & 0.143321 & 0.082686 & 0.15535 & 0.102249 & 0.179173 & 0.079576 & 0.083189 & 0.07277 & 0.085203 & 0.096694 \\
						0.426788 & 0.309989 & 0.382067 & 0.149812 & 0.223144 & 0.153431 & 0.266111 & 0.177883 & 0.338934 & 0.218452 & 0.119205 & 0.119205 & 0.177749 & 0.177749 & 0.106484 & 0.116981 & 0.144364 & 0.076263 & 0.080931 & 0.101065 \\
						0.059507 & 0.341811 & 0.228012 & 0.481689 & 0.361759 & 0.24728 & 0.405574 & 0.317089 & 0.257235 & 0.143343 & 0.108536 & 0.153521 & 0.108563 & 0.234807 & 0.111816 & 0.147381 & 0.138169 & 0.148481 & 0.116252 & 0.085719 \\
						0.041361 & 0.272979 & 0.510414 & 0.35242 & 0.382763 & 0.326018 & 0.406146 & 0.587101 & 0.522924 & 0.062991 & 0.128352 & 0.177294 & 0.12212 & 0.309878 & 0.15532 & 0.225186 & 0.058794 & 0.066287 & 0.070577 & 0.088748 \\
						0.978868 & 0.337105 & 0.381509 & 0.394685 & 0.386122 & 0.387987 & 0.388465 & 0.388278 & 0.388312 & 0.388322 & 0.388318 & 0.388318 & 0.388319 & 0.388318 & 0.388318 & 0.388318 & 0.388318 & 0.388318 & 0.388318 & 0.388318 \\
						0.967102 & 0.539468 & 0.38593 & 0.386632 & 0.38895 & 0.387838 & 0.388225 & 0.388334 & 0.388311 & 0.388317 & 0.388319 & 0.388318 & 0.388318 & 0.388318 & 0.388318 & 0.388318 & 0.388318 & 0.388318 & 0.388318 & 0.388318 \\
						0.426788 & 0.373663 & 0.403591 & 0.385442 & 0.387265 & 0.388587 & 0.388261 & 0.388297 & 0.388326 & 0.388318 & 0.388318 & 0.388319 & 0.388318 & 0.388318 & 0.388318 & 0.388318 & 0.388318 & 0.388318 & 0.388318 & 0.388318 \\
						0.059507 & 0.405793 & 0.367118 & 0.388225 & 0.390219 & 0.388086 & 0.388329 & 0.388361 & 0.388312 & 0.388318 & 0.388319 & 0.388318 & 0.388318 & 0.388318 & 0.388318 & 0.388318 & 0.388318 & 0.388318 & 0.388318 & 0.388318 \\
						0.041361 & 0.341033 & 0.431766 & 0.389178 & 0.38796 & 0.389256 & 0.388299 & 0.388289 & 0.388334 & 0.388317 & 0.388318 & 0.388319 & 0.388318 & 0.388318 & 0.388318 & 0.388318 & 0.388318 & 0.388318 & 0.388318 & 0.388318 \\
					}\datatable;
					\pgfplotstabletranspose\datatable{\datatable};
					\addplot[mark=o, blue!60!black, thick, opacity=0.2, forget plot] table [x index=1, y index=2]{\datatable};
					\addplot[mark=o, red!90!black, thick, opacity=0.2, forget plot] table [x index=1, y index=3]{\datatable};
					\addplot[mark=o, orange!80!yellow, thick, opacity=0.2, forget plot] table [x index=1, y index=4]{\datatable};
					\addplot[mark=o, green!60!black, thick, opacity=0.2, forget plot] table [x index=1, y index=5]{\datatable};
					\addplot[mark=o, cyan!90!black, thick, opacity=0.2, forget plot] table [x index=1, y index=6]{\datatable};
					\addplot[mark=o, blue!60!black, thick] table [x index=1, y index=7]{\datatable};
					\addplot[mark=o, red!90!black, thick] table [x index=1, y index=8]{\datatable};
					\addplot[mark=o, orange!80!yellow, thick] table [x index=1, y index=9]{\datatable};
					\addplot[mark=o, green!60!black, thick] table [x index=1, y index=10]{\datatable};
					\addplot[mark=o, cyan!90!black, thick] table [x index=1, y index=11]{\datatable};
					
					\nextgroupplot[title={FJ~model}]
					\pgfplotstableread[col sep=&, row sep=\\]{
						1 & 2 & 3 & 4 & 5 & 6 & 7 & 8 & 9 & 10 & 11 & 12 & 13 & 14 & 15 & 16 & 17 & 18 & 19 & 20\\
						0.977408 & 0.395437 & 0.103039 & 0.096475 & 0.100709 & 0.061842 & 0.229466 & 0.138085 & 0.098874 & 0.121901 & 0.129305 & 0.11047 & 0.147565 & 0.126073 & 0.147307 & 0.137146 & 0.105573 & 0.15387 & 0.12299 & 0.126473 \\
						0.970155 & 0.112347 & 0.451788 & 0.629878 & 0.131426 & 0.083018 & 0.154905 & 0.158315 & 0.117151 & 0.211367 & 0.314638 & 0.448993 & 0.361864 & 0.26969 & 0.326204 & 0.201302 & 0.13076 & 0.299421 & 0.220346 & 0.138564 \\
						0.197854 & 0.239697 & 0.194822 & 0.158075 & 0.093573 & 0.093573 & 0.102823 & 0.136135 & 0.128421 & 0.128421 & 0.144582 & 0.155201 & 0.155201 & 0.119976 & 0.128588 & 0.168249 & 0.210967 & 0.125885 & 0.151471 & 0.136531 \\
						0.07473 & 0.055615 & 0.194834 & 0.085997 & 0.099048 & 0.109252 & 0.109252 & 0.132165 & 0.132165 & 0.132165 & 0.132165 & 0.132165 & 0.132165 & 0.108091 & 0.113393 & 0.192877 & 0.192877 & 0.192877 & 0.186149 & 0.1973 \\
						0.051569 & 0.520554 & 0.642636 & 0.228314 & 0.331193 & 0.322219 & 0.322219 & 0.562275 & 0.50407 & 0.492513 & 0.492513 & 0.460772 & 0.497909 & 0.497909 & 0.497909 & 0.31829 & 0.095124 & 0.095124 & 0.149829 & 0.149829 \\
						0.977408 & 0.378872 & 0.203354 & 0.17464 & 0.169535 & 0.165349 & 0.160419 & 0.155218 & 0.150039 & 0.145012 & 0.140204 & 0.135648 & 0.131358 & 0.127336 & 0.123577 & 0.120072 & 0.116809 & 0.113775 & 0.110955 & 0.108336 \\
						0.970155 & 0.160689 & 0.289315 & 0.266614 & 0.23823 & 0.217688 & 0.202121 & 0.18957 & 0.179085 & 0.170115 & 0.162298 & 0.155382 & 0.149189 & 0.143593 & 0.1385 & 0.133842 & 0.129566 & 0.125628 & 0.121995 & 0.118638 \\
						0.197854 & 0.256548 & 0.204829 & 0.185316 & 0.177425 & 0.171328 & 0.16533 & 0.159417 & 0.153706 & 0.148262 & 0.143118 & 0.138284 & 0.133758 & 0.129533 & 0.125596 & 0.121932 & 0.118526 & 0.115361 & 0.112423 & 0.109696 \\
						0.07473 & 0.114165 & 0.16005 & 0.16694 & 0.163222 & 0.157914 & 0.152576 & 0.147407 & 0.142458 & 0.137761 & 0.133333 & 0.129178 & 0.125292 & 0.121667 & 0.118291 & 0.115151 & 0.112233 & 0.109522 & 0.107005 & 0.104669 \\
						0.051569 & 0.339214 & 0.343619 & 0.303124 & 0.266511 & 0.23885 & 0.217895 & 0.201537 & 0.188381 & 0.177513 & 0.168319 & 0.160379 & 0.153405 & 0.147197 & 0.141611 & 0.136545 & 0.131923 & 0.127687 & 0.123791 & 0.120201 \\
					}\datatable;
					\pgfplotstabletranspose\datatable{\datatable};
					\addplot[mark=o, blue!60!black, thick, opacity=0.2] table [x index=1, y index=2]{\datatable};\addlegendentry{$x_1(t)\quad$}
					\addplot[mark=o, red!90!black, thick, opacity=0.2] table [x index=1, y index=3]{\datatable};\addlegendentry{$x_2(t)\quad$}
					\addplot[mark=o, orange!80!yellow, thick, opacity=0.2] table [x index=1, y index=4]{\datatable};\addlegendentry{$x_3(t)\quad$}
					\addplot[mark=o, green!60!black, thick, opacity=0.2] table [x index=1, y index=5]{\datatable};\addlegendentry{$x_4(t)\quad$}
					\addplot[mark=o, cyan!90!black, thick, opacity=0.2] table [x index=1, y index=6]{\datatable};\addlegendentry{$x_5(t)$}
					\addplot[mark=o, blue!60!black, thick] table [x index=1, y index=7]{\datatable};\addlegendentry{$\widehat{x}_1(t)\quad$}
					\addplot[mark=o, red!90!black, thick] table [x index=1, y index=8]{\datatable};\addlegendentry{$\widehat{x}_2(t)\quad$}
					\addplot[mark=o, orange!80!yellow, thick] table [x index=1, y index=9]{\datatable};\addlegendentry{$\widehat{x}_3(t)\quad$}
					\addplot[mark=o, green!60!black, thick] table [x index=1, y index=10]{\datatable};\addlegendentry{$\widehat{x}_4(t)\quad$}
					\addplot[mark=o, cyan!90!black, thick] table [x index=1, y index=11]{\datatable};\addlegendentry{$\widehat{x}_5(t)$}
				\end{groupplot}
			\end{tikzpicture}
	\caption{Components of the original opinion (sentiment) vector $x(t)$ and fitted opinion vector $\widehat{x}(t)$ for heterogeneous populations of LLM agents under two opinion dynamics models. The components of the vectors correspond to the opinions of the LLM agents DeepSeek~V3, GPT-4o mini, Qwen2.5, Mistral Large, and Llama~3.3, respectively. The horizontal axis represents the round number.}
	\label{fig:heterogeneous}
\end{figure*}

Table~\ref{tab-heterogeneous} reports the results, which should be interpreted in light of the predefined initial stance parameters.  Fig.~\ref{fig:heterogeneous} visualizes the fitted opinion trajectories for a heterogeneous population of LLM agents, demonstrating a high degree of consistency with the FJ model of opinion formation. We emphasize that these results are conditioned on the particular initial stance configuration. Different initialization schemes may yield alternative patterns, especially given that LLMs are sensitive to starting positions. The analysis therefore reflects one instantiation of cross-model trust under specific initial conditions, rather than a universal characterization of model behavior.

\begin{table*}[t]
	\centering
	\footnotesize
	\caption{Self-trust and susceptibility estimates for heterogeneous LLM agents under models of social influence.}
	\label{tab-heterogeneous}
	\begin{tabular}{@{}l c *{6}{c}@{}}
		\toprule
		\multirow{2}{*}{LLM agents} 
		& \multirow{2}{*}{Model} 
		& \multicolumn{5}{c}{$\widehat{W}_{ij}$} 
		& \multirow{2}{*}{$\widehat{S}_{ii}$} \\
		\cmidrule(lr){3-7}
		&& DeepSeek~V3 & GPT-4o mini & Qwen2.5 & Mistral Large & Llama~3.3 & \\
		\midrule
		DeepSeek~V3 & DG & 0 & 0 & 0.7558 & 0.2442 & 0 & -- \\
		& FJ & 0.3369 & 0 & 0 & 0.6631 & 0 & 1 \\
				
		GPT-4o mini & DG & 0.2889 & 0.0903 & 0.3609 & 0.2523 & 0.0076 & -- \\
		& FJ & 0 & 0.0300 & 0.5500 & 0 & 0.4200 & 0.9986 \\
		
		Qwen2.5     & DG & 0.0605 & 0.1153 & 0.4191 & 0.4051 & 0 & -- \\
		& FJ & 0.1628 & 0 & 0.2884 & 0.5199 & 0.0289 & 1 \\
			
		Mistral Large & DG & 0.3327 & 0 & 0.1192 & 0.3629 & 0.1852 & -- \\
		& FJ & 0 & 0 & 0.3769 & 0.5966 & 0.0265 & 0.8613 \\
				
		Llama~3.3   & DG & 0 & 0.3152 & 0 & 0.4353 & 0.2495 & -- \\
		& FJ & 0.2746 & 0.0364 & 0 & 0 & 0.6890 & 1 \\
		\bottomrule
	\end{tabular}
\end{table*}

Combining the results from Table~\ref{tab-heterogeneous}, several consistent patterns and notable differences emerge regarding multi-agent opinion formation and trust dynamics among heterogeneous LLMs.
		
\emph{Asymmetric influence patterns.}
Across both frameworks, opinion interactions among LLM agents are highly asymmetric. Under the DG framework, DeepSeek~V3, despite its strongly positive initial opinion, is rapidly detached from its own stance and becomes primarily influenced by Qwen2.5 and Mistral Large, while receiving no influence from other agents. At the same time, DeepSeek~V3 influences GPT-4o mini and Mistral Large. GPT-4o mini affects Qwen2.5 and Llama~3.3, while itself being shaped by DeepSeek~V3, Qwen2.5, and Mistral Large. Mistral Large plays a broadly receptive role in the DG model, while influencing agents across positive, neutral, and negative stances. Llama~3.3, holding an extremely negative initial opinion, selectively influences Mistral Large, whose stance is closest to its own, while is noticeably influenced by GPT-4o mini and Mistral Large.

Under the FJ framework, these asymmetries become more pronounced and selective. DeepSeek~V3's extremely positive initial opinion is influenced by Mistral Large and its own belief, while it shapes the opinions of Qwen2.5 and Llama~3.3. GPT-4o mini is influenced by Llama~3.3 and Qwen2.5, yet only weakly influences Llama~3.3, indicating limited receptiveness. Qwen2.5, starting from a neutral position, is most strongly influenced by all agents except GPT-4o mini, while influencing GPT-4o mini and Mistral Large, reinforcing interactions among nearby viewpoints. Mistral Large is influenced by Qwen2.5 and weakly by Llama~3.3, but influences DeepSeek~V3 and Qwen2.5. Finally, Llama~3.3 is influenced by DeepSeek~V3 and weakly by GPT-4o mini, while most strongly shaping GPT-4o mini, with minor additional influence from Mistral Large and Qwen2.5.
			
\emph{Susceptibility and adaptivity.}
Differences in susceptibility across agents lead to heterogeneous responsiveness under the FJ model. Agents such as DeepSeek~V3 exhibit high adaptivity, showing strong responsiveness to external inputs and limited retention of their initial stance. In contrast, Llama~3.3 retains its extremely negative position more strongly and updates its belief primarily in response to a small subset of agents. These patterns are consistent with DG dynamics, where agents with lower effective self-influence rely more heavily on external opinions, while others exhibit stronger opinion anchoring.

\emph{Role of initial conditions.}
Initial opinion positions play a central role in shaping trust allocation and influence direction. Agents holding extreme initial stances --- DeepSeek~V3 (strongly positive) and Llama~3.3 (strongly negative) --- tend to interact selectively, avoiding agents with highly dissimilar views while assigning influence to more proximate or intermediate ones. As a result, agents with neutral or moderately positioned opinions, such as Qwen2.5 and Mistral Large, frequently mediate opinion flows between extremes. The resulting asymmetries observed in both DG and FJ influence matrices can therefore be directly linked to the initial distribution of opinions.

\emph{Model-specific differences.}
While both models capture asymmetric influence patterns, the FJ framework accentuates selectivity by explicitly incorporating susceptibility, leading to sharper differentiation in who influences whom. In contrast, the DG model emphasizes aggregate weighted averaging, producing broader and more distributed interaction patterns. As a result, mediating roles, particularly those of Qwen2.5 and Mistral Large, are expressed more diffusely in DG dynamics, whereas FJ dynamics highlight selective dependence and resistance more explicitly.

\section{Conclusion}\label{sec5}

We developed a controlled multi-agent framework to study opinion formation among LLM agents, using a neutral discussion topic to minimize prior-knowledge bias. Standardized prompts ensured consistent initial stances, enabling reproducible tracking of opinion dynamics under both homogeneous and heterogeneous settings.

In homogeneous groups, DeepSeek~V3 agents exhibited near-perfect alignment with both DG and FJ models, while Qwen2.5 showed moderate fit and Mistral Large the weakest, reflecting differences in how agents integrate peer input. Self-trust and susceptibility analyses revealed heterogeneous anchoring: Qwen2.5 strongly retained initial opinions under DG but became more adaptive under FJ, whereas Mistral Large relied heavily on peers, with the highest susceptibility. GPT-4o mini and Llama~3.3 displayed intermediate behaviors, suggesting implicit variation in aggregation mechanisms even within homogeneous populations.

In heterogeneous interactions, opinion dynamics became highly asymmetric. DG dynamics produced broad but proximity-driven influence, with extreme agents interacting selectively and intermediates such as Qwen2.5 and Mistral Large mediating between polarized viewpoints. FJ dynamics amplified these asymmetries, highlighting differences in susceptibility and selective updating rather than stable leader--follower hierarchies. Agents with high adaptivity responded rapidly to peers, whereas those with extreme stances retained more of their initial opinions.

Overall, opinion evolution among LLM agents is shaped by the interplay of initial stances, self-trust, and susceptibility. Homogeneous ensembles converge faster and more predictably, while heterogeneous systems display pronounced asymmetries and slower consensus. Classical DG and FJ models capture these dynamics systematically, providing insight into how diverse LLM architectures internalize, propagate, and resist opinions in multi-agent deliberations.

		

\section*{Declarations}

\paragraph{Conflict of interest} The authors declare that they have no conflict of interest.

\bibliographystyle{elsarticle-harv}
\bibliography{custom1}


\end{document}